# Simulating Topological Robustness of Fano Resonance in Rotated Honeycomb Photonic Crystals

J. Hajivandi[1*], E. Kaya[1], G. Edwards[2] and H. Kurt[1]

[1] Department of Electrical and Electronics Engineering

TOBB University of Economics and Technology, Ankara 06560, Turkey

[2] Department of Computer Science and Electronic & Electrical Engineering, Faculty of Science and Engineering, Thornton Science Park, University of Chester, Pool Lane, Ince, Chester, CH2 4NU, United Kingdom

*jamileh@etu.edu.tr

## Abstract

The Fano resonance with a distinctive ultra-sharp, asymmetric line shape and high quality factor Q, is a widely occurring phenomena that has a large variety of optical, plasmonic and microwave manifestations. In this paper, we explore the characteristic robustness of the Fano resonance mode, which is topologically protected by manufacturing band inversion induced by breaking the mirror symmetry of a two-dimensional honeycomb photonic crystal (HPC), associated with $C_6$ point group symmetry. So the dark and bright topological edge modes appear in the band gap made by opening of the Dirac cone. Destructive and constructive interference of the dark and bright modes leads to the asymmetric line shape of the Fano resonance. The Fano resonance mode which is very sensitive to the environmental and geometrical perturbations, can be applied to sensor design. Here we demonstrate that the topological Fano resonance mode preserves its asymmetric, ultra-sharp line shape in the presence of the disorder, defects and cavities, and this has useful optical device applications such as in low threshold lasers, and extremely precise interferometers.

Topological insulators (TIs) with their attendant insulator properties in their bulk but conduction at their surface (three dimensional) or edge (two dimensional) act as the quantum spin Hall (QSH) effect systems [1-3]. The novel behavior of the edge states including protection and robustness against disorder in the electronic systems, motivated study of analogous behavior in the photonic structures [4-5]. Moreover, the photonic topological insulators (PTIs) with the photon pseudo spin properties look to electronics systems for guidance on the manipulation of the photonic properties. The hexagonal photonic crystals (PCs), preserve the $C_6$ point group symmetry and time reversal (TR) symmetry are suitable to investigate topological insulator effects from an existence point of view. They can be used in the absence of the magnetic field set ups [6-14] to study the topological properties at the common boundary of two PCs, with different topologies. The topologically protected edge states near the Dirac cones, are scattering immune without breaking the TR symmetry by employing magnetic fields or bi-anisotropic materials [15-18]. Nowadays, there are many approaches for making modifications on the PCs to lift the Dirac point degeneracy and cause a band gap region to emerge for producing topological behavior [19-24]. Indeed, from the early findings on the Graphene models, the Dirac point was deemed important where the gradient of the dispersion bands is zero [25-26]. The honeycomb PCs have the $C_{6v}$ symmetry point group, with the four-folded degeneracies at the double Dirac cone [27-28]. By breaking the $C_{6v}$ symmetry, the bands invert with respect to each other, near the Dirac point, leading to the appearance of different topological behaviors which can be used for back-scattering immune unidirectional light propagation [29-33]. One may employ the helical edge modes, arising in the band gap region, to study the one-way light propagation at the boundary of two PCs with different trivial and non-trivial characteristics and also in waveguides made based on the topological behaviors. Moreover, one can study these features in the topological waveguides containing defects, disorders or cavities [34–43]. In our previous works [44-47], we reduced the $C_6$ symmetry of the ordinary honeycomb PC to $C_3$ point group symmetry, by modifying the distance between the rods and the center of the unit cell or by introducing small rods near main ones to remove the Dirac cone degeneracy. Removing the double Dirac cone, in the band gap at central point of the Brillion zone lead to the emergence of two PCs, with different topological behavior. These two types of PCs with simple geometry can be used for fabrication of PTIs with unique optical properties.

The interaction of discrete photonic bound states with the continuum (BICs), lead to the occurrence a Fano resonance. Indeed, such bound states in the continuum BICs have intriguing characteristics such as infinite lifetimes and Q-factor resonances which have a wide variety of application in optical techniques such as lasing and sensing. The Fano resonance may be applied to generate ultra-sharp optical pulses realizing ultra-small, low threshold lasers and beam filters [48]. From the diversity of effects in TI electronic condensed matter, PTIs are ripe for technical application opportunities, Studying the leakage of bound photonic states into free space, illuminates topological photonic phases passing through a Fano resonance [49]. The narrow line shape of topological Fano resonances shifts under environmental perturbations thus acting like a sensor which is immune to various kind of disorders and defects. The topological Fano resonances can be realized through the constructive and destructive interference of topological dark and bright edge modes with high and low quality factor (Q) respectively [50-51].

In this paper, we continue our study with the honeycomb photonic crystal (PC) that has been used in our recent works [44-45]. A two-dimensional honeycomb PC composed of six circular rods, within the unit cell, all with a relative permittivity of $\varepsilon = 11.7$, surrounded by a vacuums employed. The radius of each rods is $r$. The distance from the centre of the unit cell to the centre of the rod, as well as the distance between the adjacent rods are both $R$ (see Fig. 1). For the underlying 2D PC, the Bravais lattice vectors are $\vec{a}_1 = (a, 0)$ and $\vec{a}_1 = \left(\frac{a}{2}, \frac{a\sqrt{3}}{2}\right)$, with the lattice constant given by $a = 1\mu m$. By tuning the radii $r$ and $R$, the $C_6$ symmetry of the lattice is preserved, and multiple topological transitions are observed. The calculations in Ref. [19] show that at $R = a/3$, the sublattice symmetry of the honeycomb lattice enforces the emergence of the double Dirac point, with a four-fold degeneracy at the $\Gamma$ point of the Brillouin zone,. By breaking the inversion symmetry, the double Dirac cone will be lifted. Consequently, the topological edge state of light, with a robust pseudospin-dependent transportation will be excited.

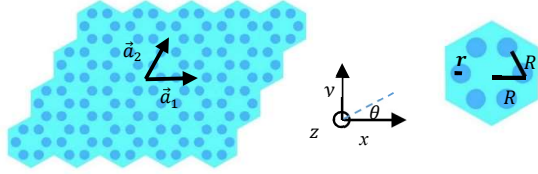

Fig 1. The schematic configuration of the honeycomb PC consists of six dielectric rods embedded in the air. $\vec{a}_1$ and $\vec{a}_2$ are the two basis vectors of the honeycomb lattice. $\theta$ is the angle that the lattice will be rotated about the $x$ axis to remove the double Dirac cone degeneracy at the $\Gamma$ point in subsequent calculations

Here we study the dispersion behavior of this PC with the parameters $R = a/3$ and $r = R/3$ for the Transverse, Magnetic Polarization, TM. We use the software electromagnetic simulation package *MIT Photonics Bands (MPB)* for calculating the band structures [53]. Numerical photonic bandstructure results plotted in the Fig. 2. (a), show the double Dirac cone has appeared at the $\Gamma$ point, due to the degeneracy between the $p$ and $d$ bands. The inset shows the electric field distributions at the double Dirac cone which consist of symmetric (S) and anti-symmetric (A) modes. The dashed lines represent the symmetric (mirror) axes. By rotating the PC about the $x$ axis by an angle of the rotation $\theta$ and consequently removing the mirror reflection symmetry of the PC, the degeneracy between the bands $p$ and $d$ is removed. Indeed, rotations of the PC lead to mismatch of the symmetry planes of the rods and lattice and breaking the reflection symmetry of the PC.

This perturbation yields an opening up of the Dirac cone and the emergence of a bandgap. Indeed, breaking the mirror symmetry by performing a rotation of the PC about the $x$ axis, opens the Dirac cone and causes a phase inversion (band inversion). Such an effect permits us to design the PCs with a robust edge transport. Fig. 2. (b). shows numerical results for the dispersion of the rotated PC for an angle $\theta = \pm 10^0$. By comparing Fig. 2 (c) and (d), we find that the Band no. 5 at $\theta = 10^0$ switched with the Band no. 4 at $\theta = -10^0$. Namely the Band no. 4 at $\theta = 10^0$ switched with the Band no. 5 at $\theta = -10^0$ and also the Band no. 3 (2) at $\theta = 10^0$ and the Band no. 2 (3) at $\theta = -10^0$ are switched to each other. Thus, when the honeycomb PC is rotated from $\theta = 10^0$ to $\theta = -10^0$, a band inversion takes place which leads to topological phase transitions. Indeed, reducing the inversion symmetry of the PC leads to the mutual coupling of the orbitals $p$ and $d$: In the other words, the upper symmetric orbital $d$ and the lower asymmetric orbital $p$, plus the upper asymmetric orbital $d$ and the lower symmetric orbital $p$. These hybridized eigen-states including the frequencies at the adjacency of the Dirac cone, follow fresh pseudospin up and down [54]. The opening then closing and reopening of the Dirac points, about the band inversion at $\theta = 0^0$, confirms that a topological phase transition has occurred and suggests a platform for studying the robust one-way transport near the topological interface. To this aim, we simulate an 'up-down' configuration composed of different PCs with rotations $\theta = -10^0$ (at the bottom) and $\theta = 10^0$ (at the top). Fig. 3 (a) indicates the supercell and the band diagram for this configuration. When the upper (down) hexagonal lattices with rotation angles $\theta = 10^0$ and $\theta = -10^0$, respectively form a photonic heterojunction, a topological edge state emerges as seen in the Fig. 3 (b). To continue, one can provide an up-down configuration with opposite PCs with rotation $\theta = -10^0$ (at the top) and $\theta = 10^0$ (at the bottom) and study the opposite mode emergence at the topological interface.

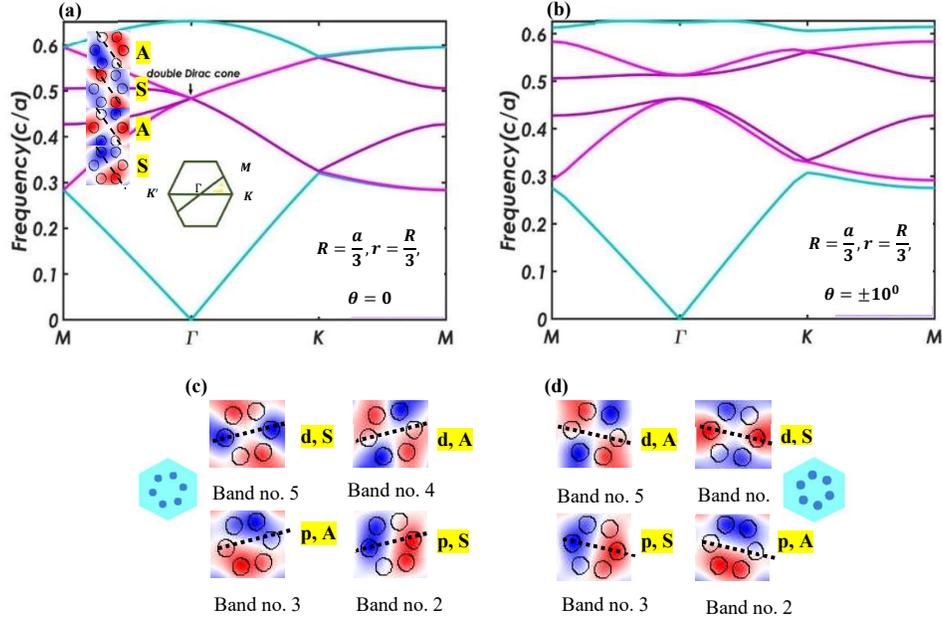

Fig. 2. (a) The band structure of the honeycomb lattice for TM polarization in which the double Dirac cone is observed at the central $\Gamma$ point of the Brillouin zone, and the symmetric and antisymmetric electric field distributions at the double Dirac cone. The symmetry axis is indicated with dashed lines. (b) The dispersion for a rotated hexagonal PC about the $x$ axis for $\theta = \pm 10^0$ with an induced bandgap, (c)-(d) the electric field, $E_z$ distributions with orbitals p and d at the $\Gamma$ point for $\theta = 10^0$ and $\theta = -10^0$ respectively.

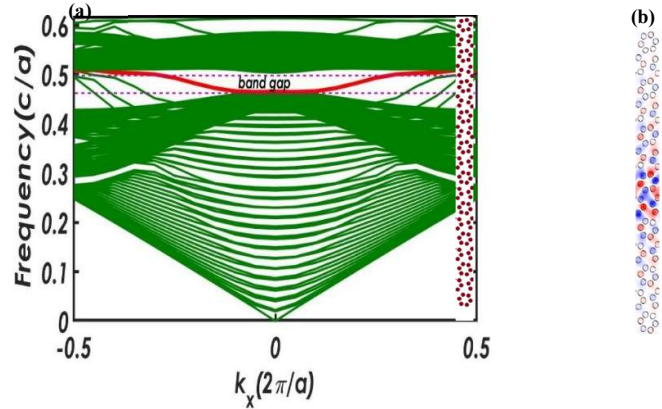

Fig. 3. (a) An up-down topological photonic supercell consisting of $25 \times 1$ PCs and related dispersion diagram and (b) with the accompanying electric field profiles at $k_x = 0$.

Another configuration is composed of right ($\theta = 10^0$) and left ($\theta = -10^0$) hexagonal unit cells with zigzag interface. As seen in the Fig. 4 (a), the helical edge states emerge within the overlapped bulk band gap of two opposite PCs which result in a pseudospin dependent unidirectional transformation. We may also study the 'reverse' configuration, consisting of right ($\theta = -10^0$) and left ($\theta = 10^0$) hexagonal unit cells which yields the similar dispersion and transformation singularity. For studying the one-way propagation of edge states, we design an up-down configuration of opposite PCs with rotation $\theta = 10^0$ and $\theta = -10^0$, Fig. 4 (b). To excite TM polarization, we use two orthogonal magnetic dipoles with 90 degree phase difference ($H_x + iH_y$), at the interface of the two PCs, to produce left hand circular polarization, counterclockwise rotation, $E_z$, which propagates at the interface. The frequency of the sources is within the range of the topological band gap. We study the robust pseudospin-dependent transportation at the $z$ interface, Fig. 4 (c). Moreover, back scattering immune propagation has been found along the interface when including disorder in the simulations. The disordered region is composed of five unit cells with inverse rotations as shown in Fig. 4 (d).

The Lumerical Finite Difference Finite Time (FDTD) software package has been used to perform the full wave simulations [55].

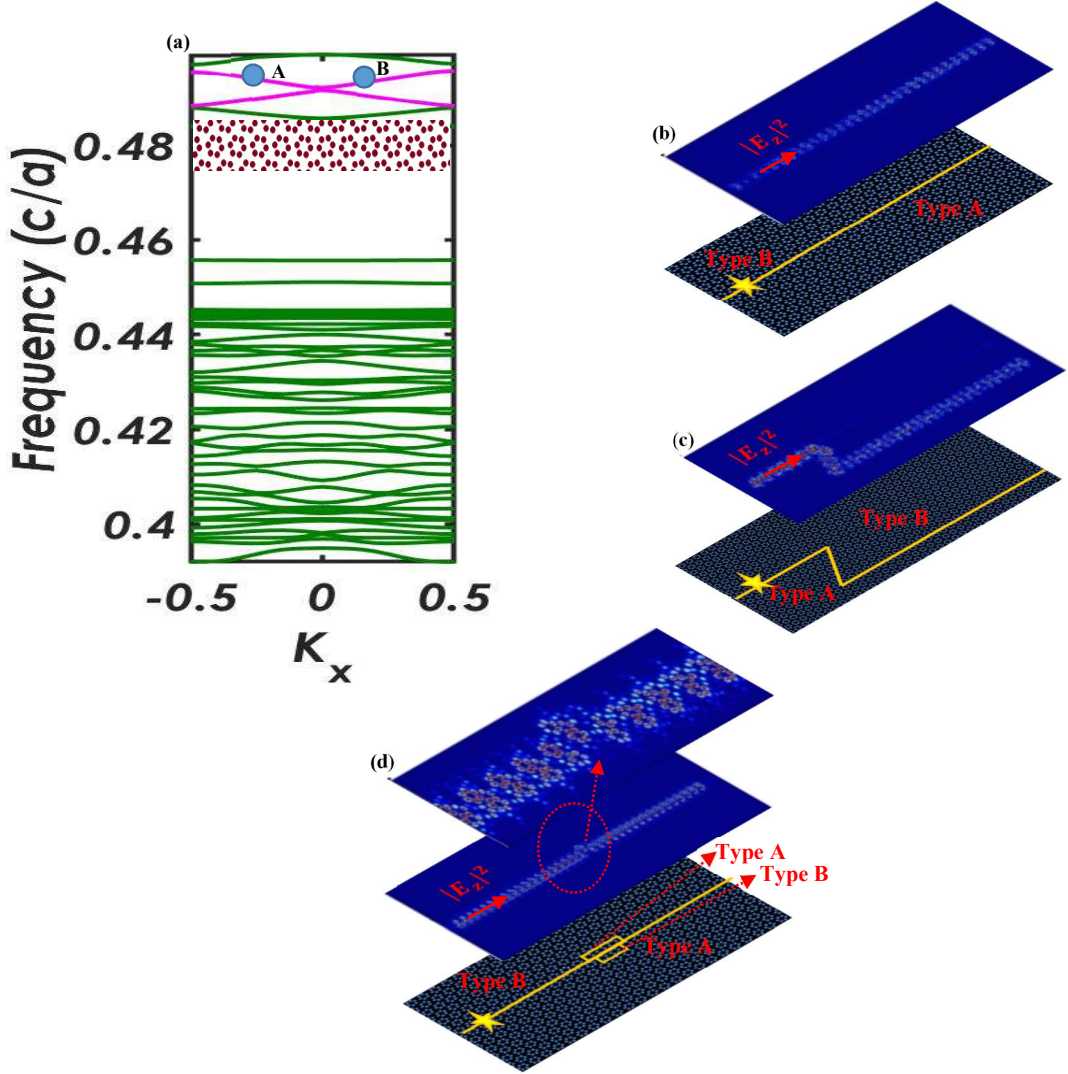

Fig. 4. (a) A right-left topological photonic supercell consisting of 13 × 2 PCs associated with related dispersion diagram, with two edge states A and B indicated on the bandstructure diagram (b) the unidirectional propagation of the electric field intensity, $|E_z|^2$, along the interface of the honeycomb PC without defect, (c) the electric field propagation around the z-shape interface of the honeycomb PC, (d) the electric field intensity transport through the topological interface including the disorder, yellow star indicates the position of the two orthogonal magnetic sources with 90 degree phase differences.

In the following, we investigate the topological Fano resonance protection in the honeycomb PTI. Considering the micro waveguide not containing any rods, Fig. 5. (a) . Thanks to the vertical inversion symmetry respecting to the central line, there will be corresponding even and odd eigenstates. We can understand that in spite of the even wave, the odd wave keeps a cutoff frequency, $f'' = \pi c/2h$ (where in $c$ indicate the speed of light and $h$ is the distance between the center of the unitcell and a boundary) by solving the scalar Helmholtz equation applying the Neumann along the y direction [52].

Next, consider the waveguide including six rods in each unit cell, Fig. 5. (a). Regarding the position of the dielectric rods at the center of the unit-cell, they can support even and odd eigenstates of $E_z$ due to the vertical symmetry of the unit-cell. Coexisting the odd waveguide mode resonance frequency in the even modes spectral range, may be happen localizing it below the $f''$. Because of different symmetry of the odd and even modes, they are bounding purely to the rods, instantaneously [51, 52]. Fig. 5. (b) illustrates the band structure and profiles of the $z$ component of the electric field, $E_z$ of the equivalent bound-state in the continuum (BIC) mode for a rhombic unit cell containing six rods rotating $0^0$, $10^0(-10^0)$ with the periodic boundary conditions (B. Cs.) along the $x$ axis and the Perfect Electrical Conductors (PECs) at the other sides. As seen from the Fig. 5 (b), the bands which are degenerate for the rotation $0^0$, are opened for the rotation $10^0(-10^0)$. Besides, the electric field profiles at the $\Gamma$ point are symmetric and antisymmetric. Indeed, the mutual coupling of the upper symmetric orbital $d$ and the lower asymmetric orbital $p$, plus the upper asymmetric orbital $d$ and the lower symmetric orbital $p$ has been detected. These hybridized eigen-states including the frequencies at the adjacency of the Dirac cone, follow fresh pseudospin up and down [54].

We apply the antisymmetric and symmetric B. Cs. to construct the odd-symmetric (dark) and even-symmetric (bright) modes. By the coupling of two resonant bright and dark modes with different lifetimes (high (low) Q or equivalently narrow (wide) band transmission respectively), then destructive and constructive interference between them, the Fano resonance mode with a typical anti-symmetric line shape is routinely realized [50].

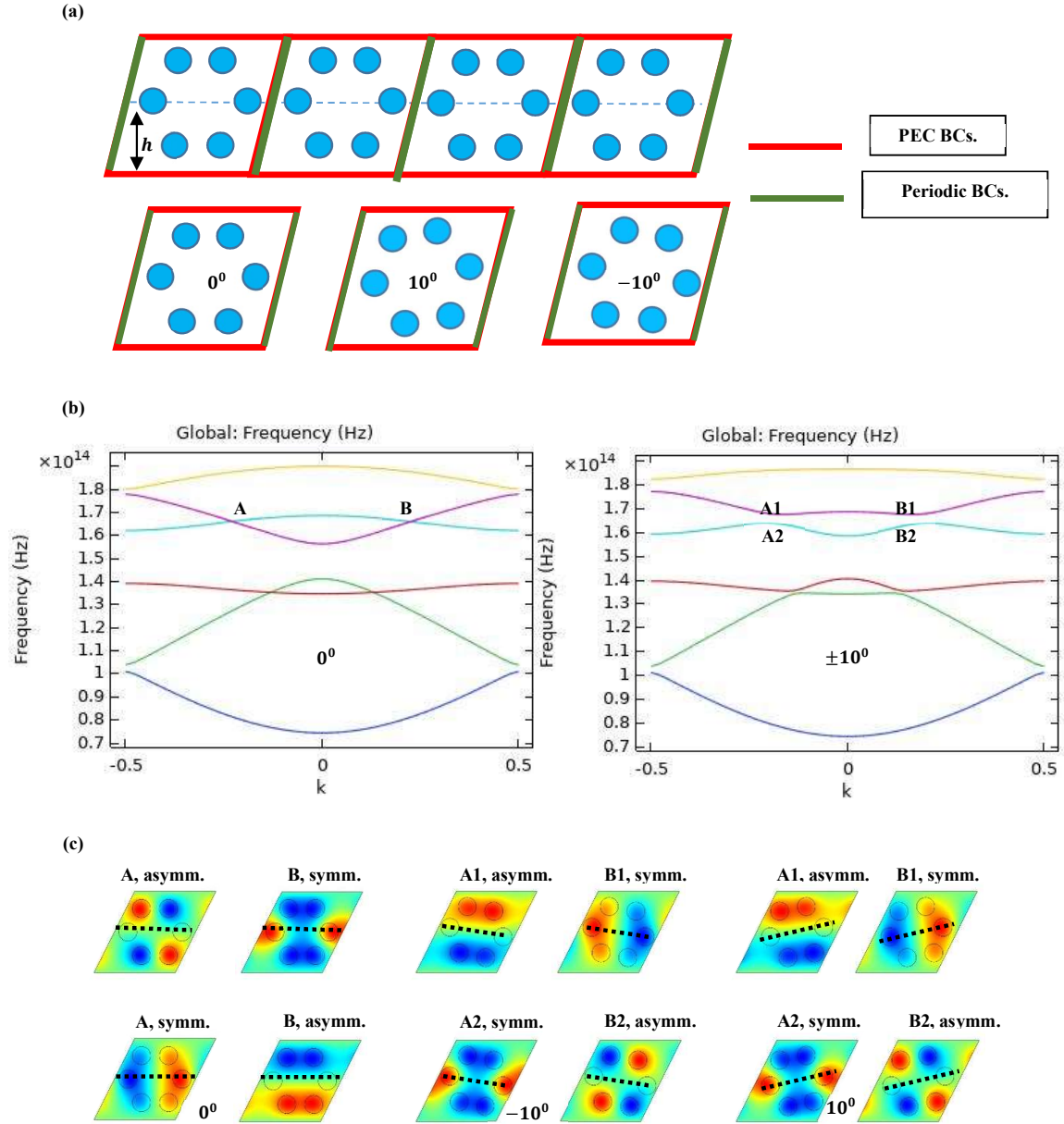

Fig. 5. (a) The microwave waveguide including six rods in each unit cell with PEC and periodic BCs., (b) The band structure and profiles of the z component of the electric field, $E_z$ of the equivalent bound-state in the continuum (BIC) mode for a hexagonal unit cell containing six rods rotating $0^0, -10^0(10^0)$ with the periodic BCs. along the lattice vectors, $\vec{a}_1$ and the Perfect Electrical Conductors (PECs) BCs at the other sides. The bands which are degenerate for the rotation angle $0^0$, are opened for the rotation angle $10^0(-10^0)$. The electric field profiles for the k=0 point are symmetric and antisymmetric and can be converted to each other (the dashed lines are the symmetric axis).

Owing to the high Q characterization, the conventional Fano resonance mode which is created by interaction of just two odd and even modes, is not preserved under perturbations and deformations of the geometry, losing the simple asymmetric line shape. So, a small defect or the presence of the disorder leads to the shifting of the resonance frequency, and deforming the Fano resonance line shape. However, the topological Fano resonance which is studied in the following, is preserved under deformation due to the dark and bright edge states being protected against perturbations [50].

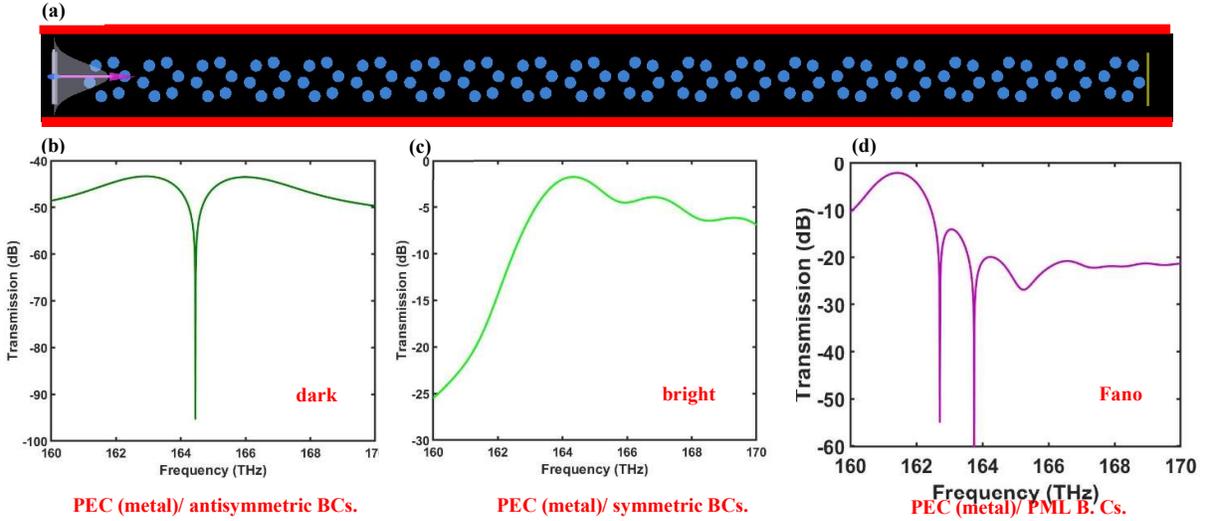

Fig. 6. (a) A topological structure composed of trivial and non-trivial PCs (rotations $10^0$ and $-10^0$) and the transmission against frequency of this structure for (b) antisymmetric (c) symmetric (d) PEC BCs. : exciting both dark and bright modes, so coupling and interfacing them, lead to creating the Fano resonance mode with ultrasharp, antisymmetric shape.

Consider a rhombic supercell composed of trivial and non-trivial PCs rotating around the angle $10^0$ and $-10^0$, and applying symmetric or anti-symmetric PECs boundary conditions along the y direction. Then, the topological Fano resonance mode will be created by the interference between the bright and dark edge modes constructively and destructively, see Fig. 6.

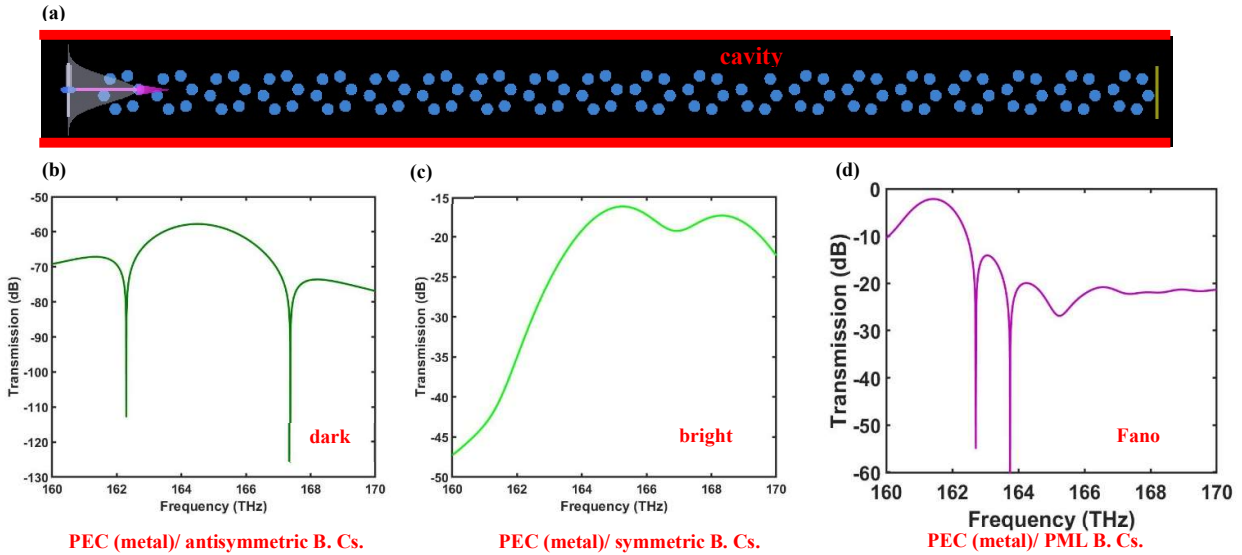

Fig. 7. (a) A topological structure composed of trivial and non-trivial PCs (rotations $10^0$ and $-10^0$) including a cavity and the transmission against frequency of this structure for (b) antisymmetric (c) symmetric (d) PEC BCs. : the Fano resonance shape has been preserved in topological composition including disorder.

Applying the Gaussian source with the topological edge frequency at the left side of the structure and the symmetric/anti-symmetric BCs. applied along the y direction, we study the transmission of the bright and dark modes. The anti-symmetric shape transmission of Fano mode will be emerged by applying just one of the PECs. The dark (anti-symmetric) and bright (symmetric) edge modes may be exited applying a single PEC to the top or bottom side. So the constructive and destructive interferences of them generate the topological Fano resonance mode. The Fano resonance mode will be topologically protected in the vicinity of the defects, disorder and even against introducing a cavity. For studying the robustness of the topological Fano mode, we prepared a topological photonic structure including various types of disorders and cavity. Figs 7 -11 show the transmission diagrams of dark/ bright and Fano modes for the topological structures including various defects. For example, in Fig. 7., we create a cavity at the topological structure by removing one of the rods and demonstrate that the topological Fano resonance shape has been protected in the vicinity of the cavity.

In order to study the topological Fano resonance protection in the vicinity of the defects, we modify the distances of the six rods from the center of their unit cells, three of them by $\Delta R = -5.348 nm$ and the other three ones by $\Delta R = -3.231 nm$ . As seen in the Fig. 8., the shape of the Fano resonance mode is preserved against the disorder (Disorder Case 1).

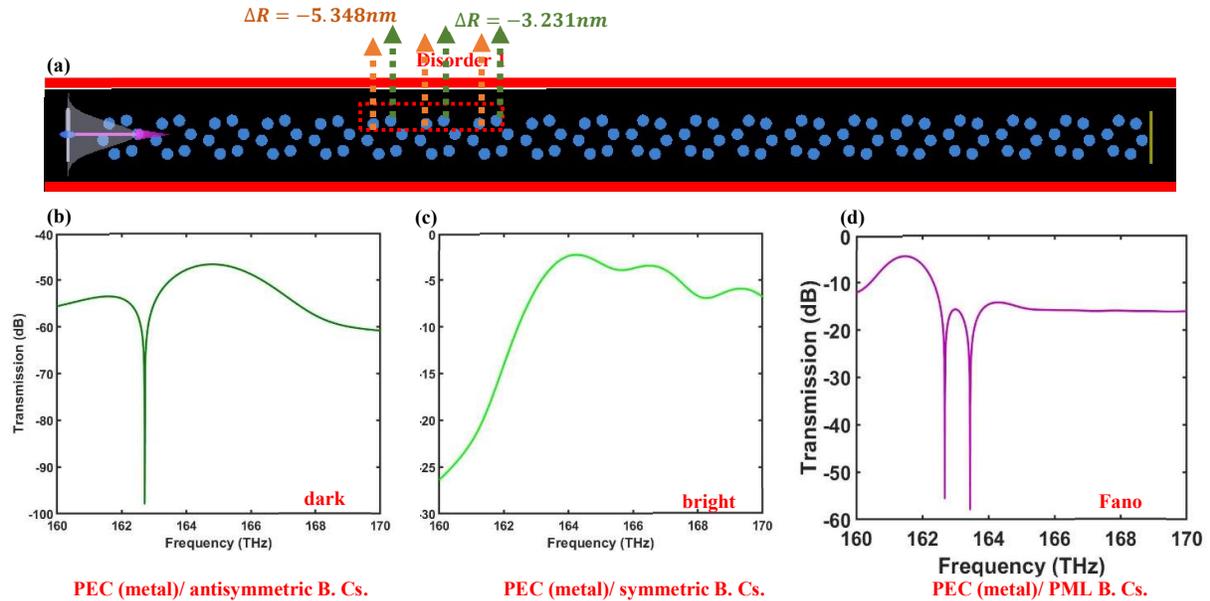

Fig. 8. (a) A topological structure composed of trivial and non-trivial PCs (rotations $10^0$ and $-10^0$) including a disorder region (indicated with a red rectangle region) and the transmission against frequency of this structure for (b) antisymmetric (c) symmetric (d) PEC BCs. : the Fano resonance shape has been preserved in topological composition including disorder.

For studying the robustness of the Fano resonance transmission through the disorder topological structure, we modify the radii of the indicated rods in the Fig. 9., by $\Delta r = -1.1 nm$, (Disorder Case 2).

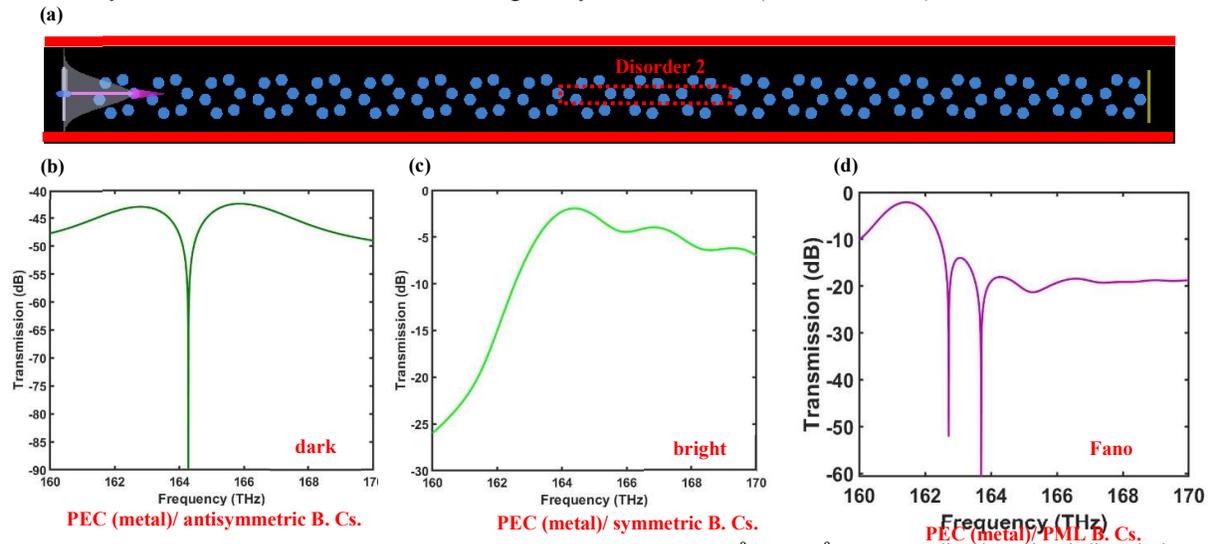

Fig. 9. (a) A topological structure composed of trivial and non-trivial PCs (rotations $10^0$ and $-10^0$) including a disorder region (indicated with a red rectangle region) and the transmission against frequency of this structure for (b) antisymmetric (c) symmetric (d) PEC BCs. : the Fano resonance shape has been preserved in topological composition including disorder.

In Fig. 10., in the disorder region indicated by a red square, the rod has been rotated to an angle of $-15^0$ instead of $10^0$. Again the Fano resonance mode protection clearly demonstrated in the numerical results, (Disorder Case 3).

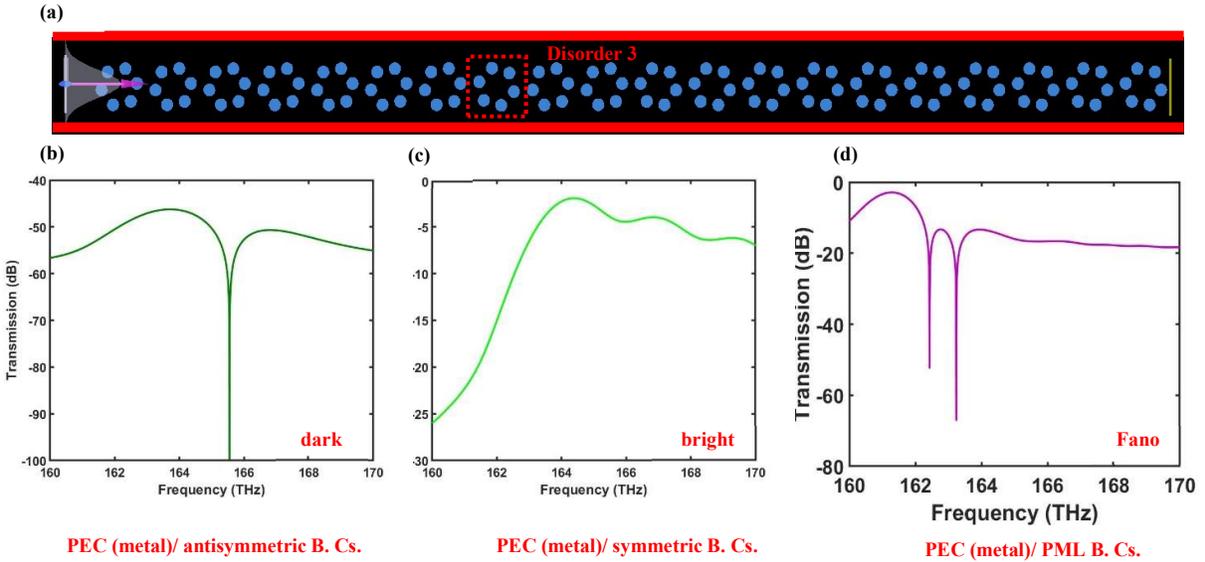

Fig. 10. (a) (a) A topological structure composed of trivial and non-trivial PCs (rotations $10^0$ and $-10^0$) including a disorder region (indicated with a red square region) and the transmission against frequency of this structure for (b) antisymmetric (c) symmetric (d) PEC BCs. : the Fano resonance shape has been preserved in topological composition including disorder.

In Fig. 11., the permittivity of the indicated unit cell, has been changed to $\varepsilon = 11.22$ (Disorder Case 4). For comparison of dark, bright and Fano resonance mode behavior through the various types of defects and the presence of a cavity, we combine the results of the Fig. 7 to Fig. 11 together in Fig. 12. As seen in the Fig. 12., although there is some Fano resonance position shifting in the various disordered topological structures, the antisymmetric shape of the Fano resonance state has been categorically preserved.

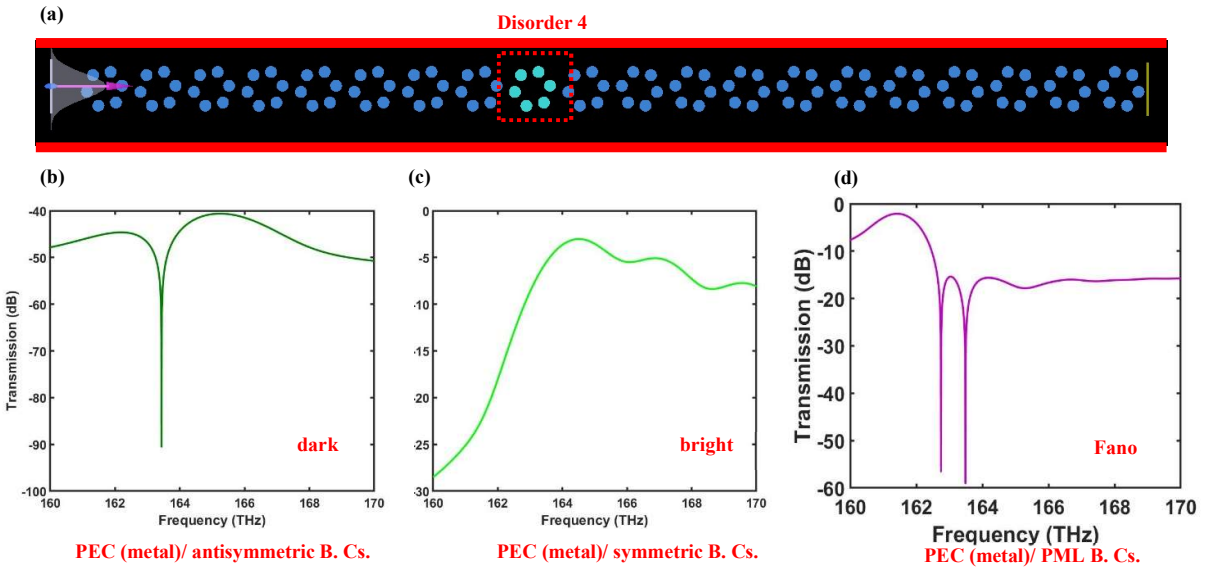

Fig. 11. (a) A topological structure composed of trivial and non-trivial PCs (rotations $10^0$ and $-10^0$) including a disorder region (indicated with a red square region) and the transmission against frequency of this structure for (b) antisymmetric (c) symmetric (d) PEC BCs. : the Fano resonance shape has been preserved in topological composition including disorder.

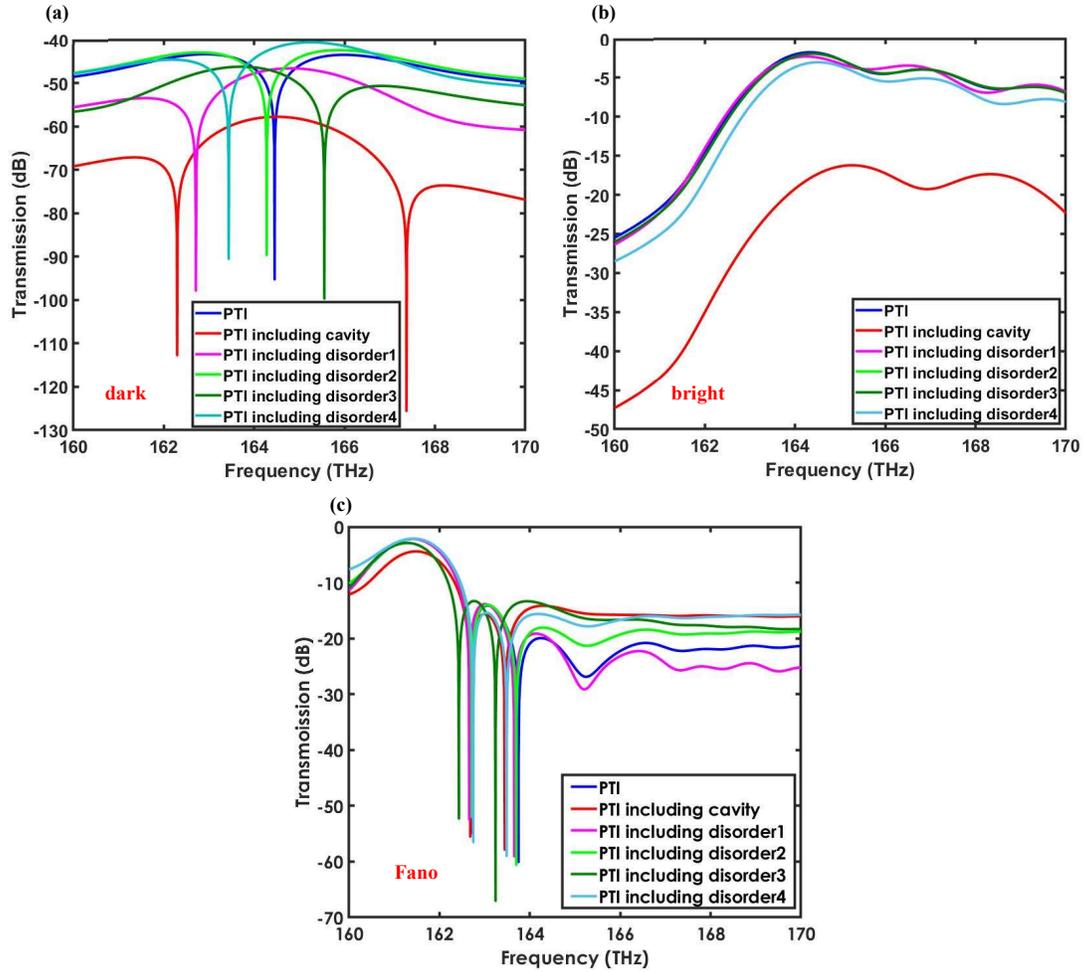

Fig. 12. Comparison of the transmission against frequency of (a) dark (b) bright and (c) Fano resonance mode for a topological structure including cavity, and disorder cases (1, 2, 3, 4). The Fano resonance mode shape is preserved for the different cavity and disorder cases. The shifting Fano resonance frequency position, suggests that this device is suitable for optical sensor applications.

For studying the Fano resonance mode in trivial structure composed of PCs with a rotation angle of zero, just as for the earlier topological structures, we apply one PEC BCs. For the trivial structure, a mode similar to the Fano resonance mode appears in our simulations which is not preserved when a cavity or defects are introduced. For introducing the disorder in the trivial structure, in Fig. 13, we modify the distances of the six rods from the center of their unit cells, with three of them given by $\Delta R = -5.348 nm$ and the other three given by $\Delta R = -3.231 nm$.

**Conclusion**

The Fano resonance mode with a characteristic ultra-sharp, asymmetric line shape is widely encountered in optics and has many applications in lasing, sensing and switching optical devices. The asymmetric shape is produced as a consequence of constructive and destructive interference of dark and bright modes, due to interaction of the bound modes and the continuum of propagating waves. We numerically investigate the individual robust Fano resonance mode which is topologically protected by the band inversion, induced by breaking the mirror symmetry of a two-dimensional honeycomb photonic crystal associated with $C_6$ point group symmetry. In contrast to the general Fano resonance mode which is very sensitive to environmental and geometrical perturbations, the topological Fano resonance mode can be realized through the interaction of topological dark and bright edge modes with a high quality factor, Q and low Q respectively and preserves ultra-sharp, asymmetric line shape in the vicinity of the cavity and defects.

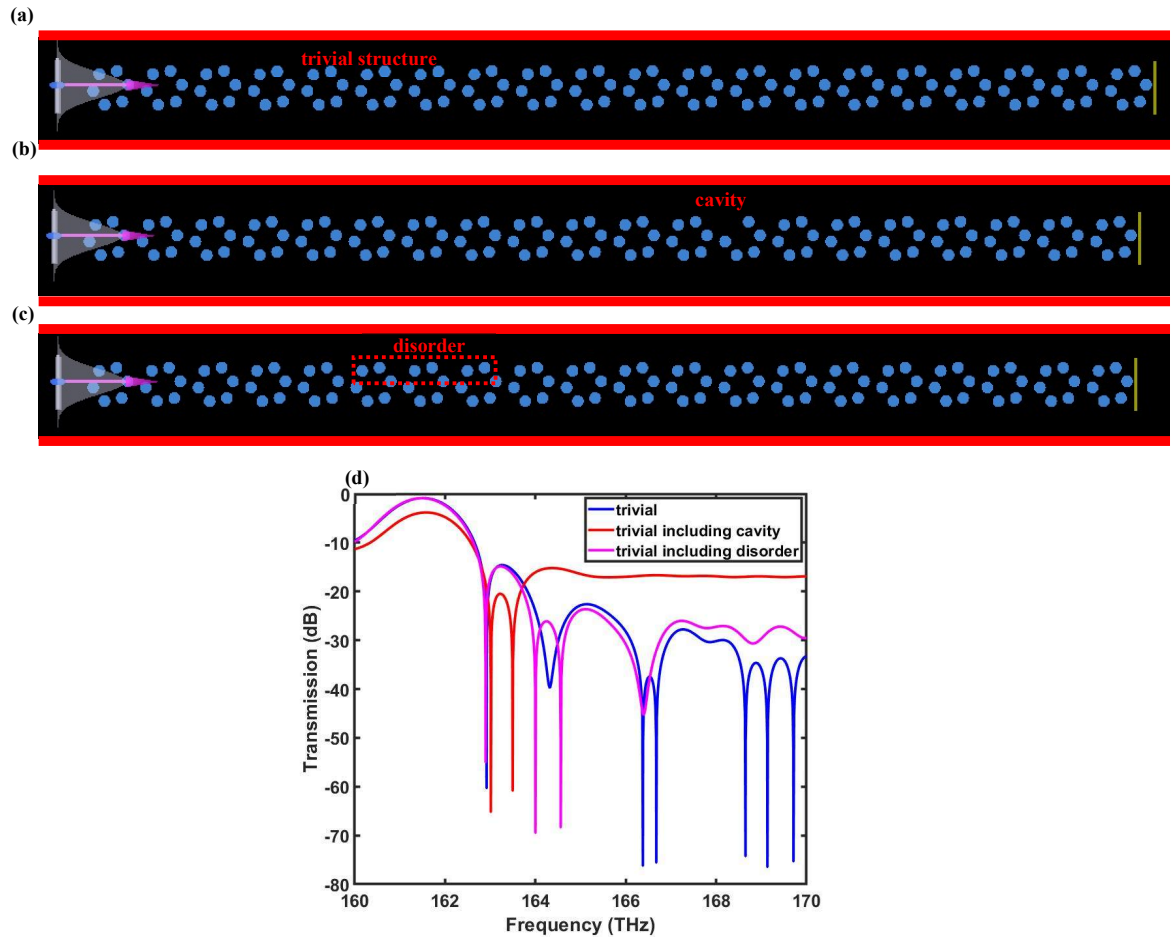

Fig. 13. (a) The structure composed of trivial PCs with PEC BCs. including (b) a cavity region, (c) disorder region, (d) Comparison of the transmission against frequency for the structures part (a)-(c), the shape of the generated mode is not preserved through the different cavity and disorder.

**Acknowledgment**

H. K. acknowledges partial support of the Turkish Academy of Sciences.

**References**

[1] C. L. Kane and E. J. Mele, "Quantum spin Hall effect in Graphene", Phys. Rev. Lett. **95**, 226801 (2005).
[2] B. A. Bernevig and S. C. Zhang, "Quantum spin Hall effect", Phys. Rev. Lett. **96**, 106802 (2006).
[3] M. Konig, S. Wiedmann and C. Brune et al., "Quantum spin hall insulator state in HgTe quantum wells", Science **318**, 766–770 (2007).
[4] Y. Kang, X. Ni and X. Cheng, et al., "Pseudo-spin–valley coupled edge states in a photonic topological insulator", Nat. Commun. **9**, 3029 (2018).
[5] L. He, Y. f. Gao and Z. Jiang, et al., "A unidirectional air waveguide basing on coupling of two self-guiding edge modes", Opt. Laser Technol. **108**, 265–272 (2018).
[6] T. Ozawa, H. M. Price, A. Amo, N. Goldman, M. Hafezi, L. Lu, M. C. Rechtsman, D. Schuster, J. Simon, O. Zilberberg, I. Carusotto, "Topological photonics", Rev. Mod. Phys. **91**, 015006 (2019).
[7] M. S. Rider, S. J. Palmer, S. R. Pocock, X. Xiao, P. Arroyo Huidobro, V. Giannini, "A perspective on topological nanophotonics: Current status and future challenges", J. Appl. Phys. **125**, 120901 (2019).

[8] M. G. Silveirinha, "Proof of the bulk-edge correspondence through a link between topological photonics and fluctuation-electrodynamics", Phys. Rev. X **9**, 011037 (2019).
[9] B. Orazbayev, N. Kaina, R. Fleury, "Chiral waveguides for robust waveguiding at the deep subwavelength scale", Phys. Rev. Appl. **10**, 054069 (2018).
[10] B. Orazbayev, R. Fleury, Nanophononics "Quantitative robustness analysis of topological edge modes in C6 and valley-Hall metamaterial waveguides", Nanophotonics **8**, 1433–1441(2019).
[11] Z. Jiang, Y. f. Gao, L. He, J. P. Sun, H. Song, Q. Wang, "Manipulation of pseudo-spin guiding and flat bands for topological edge states", Phys. Chem. Chem. Phys. **21**, 11367–11375 (2019).
[12] R. Chaunsali, W. Chen, C. J. Yang, "Subwavelength and directional control of flexural waves in zone-folding induced topological plates", Phys. Rev. B **97**, 054307. 24 (2018).
[13] S. Yves, R. Fleury, T. Berthelot, M. Fink, F. Lemoult, G. Lerosey, "Crystalline metamaterials for topological properties at subwavelength scales", Nat. Commun. **8**, 16023 (2017).
[14] S. Yves, R. Fleury, F. Lemoult, M. Fink, G. Lerosey, "Topological acoustic polaritons: robust sound manipulation at the subwavelength scale", New J. Phys. **19**, 075003 (2017).
[15] L. J. Maczewsky, J. M. Zeuner and S. Nolte, et al., "Observation of photonic anomalous Floquet topological insulators", Nat. Commun. **8**, 13756 (2017).
[16] Y. Yang, Y. F. Xu and T. Xu, et al., "Visualization of a unidirectional electromagnetic waveguide using topological photonic crystals made of dielectric materials", Phys. Rev. Lett. **120**, 217401 (2016).
[17] S. Kruk, A. Slobozhanyuk and D. Denkova et al., "Edge States and Topological Phase Transitions in Chains of Dielectric Nanoparticles", Small **13**, 1603190 (2017).
[18] P. Lodahl, S. Mahmoodian, and S. Stobbe, "Interfacing single photons and single quantum dots with photonic nanostructures", Rev. Mod. Phys. **87**, 347 (2015).
[19] L. H. Wu and X. Hu, "Scheme for achieving a topological photonic crystal by using dielectric material", Phys. Rev. Lett. **114**, 223901 (2015).
[20] T. Ma, A. B. Khanikaev, S. H. Mousavi, and G. Shvets, "Guiding electromagnetic waves around sharp corners: Topologically protected photonic transport in metawaveguides", Phys. Rev. Lett. **114**, 127401 (2015).
[21] T. Ma and G. Shvets, "All-Si Valley-Hall photonic topological insulator", New. J. Phys. **18**, 025012 (2016).
[22] L. Xu, H. X. Wang, Y. D. Xu, H. Y. Chen, and J. H. Jiang, "Accidental degeneracy in photonic bands and topological phase transitions in two-dimensional core-shell dielectric photonic crystals", Opt. Express **24**, 18059 (2016).
[23] X. D. Chen and J. W. Dong, "Valley-protected backscattering suppression in silicon photonic graphene", arXiv:1602.03352 (2016).
[24] S. Barik, H. Miyake, W. DeGottardi, E. Waks, M. Hafezi, "Chiral quantum optics using a topological resonator", New J. Phys. **18**, 113013 (2019).
[25] Y. Zhang, Y. W. Tan, H. L. Stormer and P. Kim, "Experimental observation of the quantum Hall effect and Berry's phase in graphene", Nature **438**, 201 (2005).
[26] H. Dai, T. Liu, J. Jiao, B. Xia and D. Yu, "Double Dirac cone in two-dimensional phononic crystals beyond circular cells", J. Appl. Phys. **121**, 135105 (2017).
[27] Y. Li, Y. Wu, and J. Mei, "Double Dirac cones in phononic crystals", Appl. Phys. Lett. **105**, 014107 (2014).
[28] S.Y. Yu, Q. Wang, L.Y. Zheng, C. He, X. P. Liu, M. H. Lu, and Y. F. Chen, "Acoustic phase-reconstruction near the Dirac point of a triangular phononic crystal", Appl. Phys. Lett. **106**, 151906 (2015).
[29] C. He, X. Ni, H. Ge, X. C. Sun, Y. B. Chen, M. H. Lu, X. P. Liu, L. Feng, and Y. F. Chen, "Acoustic topological insulator and robust one-way sound transport", Nat. Phys. **12**, 1124 (2016).
[30] Z. W. Zhang, Q. Wei, Y. Cheng, T. Zhang, D. J. Wu, and X. J. Liu, "Topological Creation of Acoustic Pseudospin Multipoles in a Flow-Free Symmetry-Broken Metamaterial Lattice", Phys. Rev. Lett. **118**, 084303 (2017).
[31] J. Y. Lu, C. Y. Qiu, S. J. Xu, Y. T. Ye, M. Z. Ke, and Z. Y. Liu, "Dirac cones in two-dimensional artificial crystals for classical waves", Phys. Rev. B **89**, 134302 (2014).
[32] J. Mei, Y. Wu, C. T. Chan, and Z. Q. Zhang, "First-principles study of Dirac and Dirac-like cones in phononic and photonic crystals", Phys. Rev. B **86**, 035141 (2012).
[33] H Dai, T Liu, J Jiao, B Xia, D Yu, "Double Dirac cone in two-dimensional phononic crystals beyond circular cells", J. Appl. Phys. **121**, 135105 (2017).
[34] E. Waks and J. Vuckovic, "Dipole induced transparency in Drop-filter cavity-waveguide systems", Phys. Rev. Lett. **96**, 153601 (2006).
[35] P. R. Villeneuve, D. S. Abrams, S. Fan, and J. Joannopoulos, "Single-mode waveguide microcavity for fast optical switching", Opt. Lett. **21**, 2017 (1996).
[36] S. Fan, P. R. Villeneuve, J. D. Joannopoulos, and H. A. Haus, "Channel drop filters in photonic crystals", Opt. Exp. **3**, 4 (1998).

[37] S. Fan, "Sharp asymmetric line shapes in side-coupled waveguide-cavity systems", Appl. Phys. Lett. **80**, 908 (2002).
[38] Z. Wang and S. Fan, "Compact all-pass filters in photonic crystals as the building block for high-capacity optical delay lines", Phys. Rev. E **68**, 066616 (2003).
[39] K. Nozaki, T. Tanabe, A. Shinya, S. Matsuo, T. Sato, H. Taniyama, and M. Notomi, "Sub-femtojoule all-optical switching using a photonic-crystal nanocavity", Nat. Photonics **4**, 477 (2010).
[40] G. Dong, Y. Zhang, J. F. Donegan, B. Zou, and Y. Song, "Multi-Band-Stop Filter for Single-Photon Transport Based on a One-Dimensional Waveguide Side Coupled with Optical Cavities", Plasmonics **9**, 1085 (2014).
[41] Q. Hu, B. Zou, and Y. Zhang, "Transmission and correlation of a two-photon pulse in a one-dimensional waveguide coupled with quantum emitters", Phys. Rev. A **97**, 033847 (2018).
[42] Q. Jiang, Q. Hu, B. Zou, and Y. Zhang, "Single microwave photon switch controlled by an external electrostatic field", Phys. Rev. A **98**, 023830 (2018).
[43] J. C. Yin, L. G. B. Liu et al., "Transport tuning of photonic topological edge states by optical cavities", Phys. Rev. A **99**, 043801 (2019).
[44] J. Hajivandi and H. Kurt, "Robust transport of the edge modes along the photonic topological interfaces of different configurations", arXiv:2005.08775 [physics.app-ph] (2020).
[45] J. Hajivandi and H. Kurt, "Topological photonic states and directional emission of the light exiting from the photonic topological structure composed of two dimensional honeycomb photonic crystals with different point group symmetries", arXiv:2002.11979 [physics.app-ph] (2020).
[46] J. Hajivandi and H. Kurt, "All-dielectric two-dimensional modified honeycomb lattices for topological photonic insulator and various light manipulation examples in waveguide, cavity and resonator", arXiv:2002.00588 [physics.app-ph] (2020).
[47] J. Hajivandi and H. Kurt, "Topological phase transition of the centered rectangular photonic lattice", arXiv:2005.11916 [physics.app-ph] (2020).
[48] D. R. Abujetas, et. al., "Spectral and temporal evidence of robust photonic bound states in the continuum on terahertz metasurfaces", Optica **6**, 8 (2019).
[49] M. A. Gorlach, X. Nil, et. al., "Far-field probing of leaky topological states in all-dielectric metasurfaces", Nat. Commun. **9**, 909 (2018).
[50] W. Wang, Y. Jin, et. al., "Robust Fano resonance in a topological mechanical beam", Phys. Rev. B **101**, 024101 (2020).
[51] F. Zangeneh Nejad and R. Fleury, "Topological Fano Resonances", Phys. Rev. Lett. **122**, 014301 (2019).
[52] C. W. Hsu, B. Zhen, A. D. Stone, J. D. Joannopoulos, and M. Soljacic, "Bound states in the continuum", Nat. Rev. Mater. **1**, (2016).
[53] S. G. Johnson and J. D. Joannopoulos, "Block-iterative frequency-domain methods for Maxwell's equations in a planewave basis", Opt. Exp. **8**, 173 (2001).
[54] B. Z. Xia, T. T. Liu et al., "Topological phononic insulator with robust pseudospin-dependent transport", Phys. Rev. B **96**, 094106 (2017).
[55] Lumerical Inc. http://www.lumerical.com/tcad-products/fdtd/